\documentclass[]{spie}  

\usepackage{amsmath,amsfonts,amssymb}
\usepackage{graphicx}
\usepackage[colorlinks=true, allcolors=blue]{hyperref}

\usepackage{amsfonts}
\usepackage{amsmath}
\usepackage{graphicx}      
\usepackage{float} 
\usepackage{psfrag} 
\usepackage{mathtools}
\usepackage{url}
\usepackage{cite}

\title{Subspace identification of low-dimensional Structural-Thermal-Optical-Performance (STOP) models of reflective optics}

\author[a]{Aleksandar Haber}
\author[b]{John E. Draganov}
\author[b]{Michael Krainak}

\affil[a]{Department of Manufacturing and Mechanical Engineering Technology, College of Engineering Technology, Rochester Institute of Technology,  70 Lomb Memorial Dr., Rochester, NY 14623, USA}
\affil[b]{Relative Dynamics, Inc., 6401 Golden Triangle Dr. STE 201, Greenbelt, MD 20770, USA}

\authorinfo{Further author information: (Send correspondence to A. H.)\\A. H.: E-mail: aleksandar.haber@gmail.com, Telephone: +1 585-475-5813}

\begin{document} 
\maketitle

\begin{abstract}
In this paper, we investigate the feasibility of using subspace system identification techniques for estimating transient Structural-Thermal-Optical Performance (STOP) models of reflective optics. As a test case, we use a Newtonian telescope structure. This work is motivated by the need for the development of model-based data-driven techniques for prediction, estimation, and control of thermal effects and thermally-induced wavefront aberrations in optical systems, such as ground and space telescopes, optical instruments operating in harsh environments, optical lithography machines, and optical components of high-power laser systems. We estimate and validate a state-space model of a transient STOP dynamics. First, we model the system in COMSOL Multiphysics. Then, we use LiveLink for MATLAB software module to export the wavefront aberrations data from COMSOL to MATLAB. This data is used to test the subspace identification method that is implemented in Python. One of the main challenges in modeling and estimation of STOP models is that they are inherently large-dimensional. The large-scale nature of STOP models originates from the coupling of optical, thermal, and structural phenomena and physical processes. Our results show that large-dimensional STOP dynamics of the considered optical system can be accurately estimated by low-dimensional state-space models. Due to their low-dimensional nature and state-space forms, these models can effectively be used for the prediction, estimation, and control of thermally-induced wavefront aberrations. The developed MATLAB, COMSOL, and Python codes are available online.
\end{abstract}

\keywords{Adaptive optics, structural-thermal-optical-performance (STOP) models, system identification, model-based control, telescopes}

\section{INTRODUCTION}
\label{sec:intro}  

Thermally-induced mechanical deformations, wavefront aberrations, and large focal shifts can negatively affect performance and significantly limit the resolution of both refractive and reflective optical systems. For example, thermal phenomena and thermally-induced aberrations can limit the achievable resolution and performance of optical lithography systems~\cite{choi2013lens,ravensbergen2013deformable,haber2013predictive,zhao2018active,habets2016multi,bikcora2014thermal,heertjes2020control,haber2013identification,bikcora2016parameter,bikcora2012lens},
space and ground telescopes~\cite{yoder2017opto,holzlohner2022structural,brooks2022precision,havey2019challenges,segato2011method,brooks2017predictive,zhang2020optimization,stahl2020advanced,stahl2020predictive,banyal2013opto,buleri2019structural,blaurock2005structural,gu2019thermal},
gravitational wave detectors~\cite{loriette2003absorption,zhao2006compensation,ramette2016analytical}, high power lasers~\cite{lyu2021stop,schmidt2019energy,abt2008temporal}, and other optical systems~\cite{turella2019structural,koppen2018topology,nordera2021methodology,li2022multilayer}. In the case of refractive optical systems consisting of lenses, absorbed thermal energy and non-uniform temperature distributions across optical elements, induce mechanical deformations and variations of refractive indices. These effects can in turn induce large focal shifts and wavefront aberrations. On the other hand, in the case of reflective optical elements, thermally created mechanical deformations are the main cause of thermally-induced wavefront aberrations. Here it should be noted that even if all internal optical elements are properly thermally insulated, thermally induced deformations of enclosures, supports, and other devices that are in direct mechanical contact with optics can cause significant optical misalignments. 
 
To design effective control strategies for the compensation of thermally-induced wavefront aberrations or to design novel wavefront correction devices that are based on thermo-mechanical actuation, it is often necessary to develop high-fidelity models of thermally-induced mechanical deformations and wavefront aberrations. Apart from this, high-fidelity models are important for performance prediction and worst-case analysis of optical systems under the negative influence of thermal effects. To model thermally-induced wavefront aberrations it is necessary to couple structural and thermal partial differential equations with optical parameters and ray propagation equations. These models are often referred to as Structural-Thermal-Optical-Performance (STOP) models. The development of accurate STOP models is a challenging task. First of all, STOP models involve different time scales of physical processes, as well as different types of partial differential equations and boundary conditions. Consequently, STOP models can often be numerically stiff and difficult for discretization and simulation. Secondly, for the development of efficient prediction and control algorithms, it is crucial to obtain low-dimensional models. However, discretized STOP models obtained by applying finite-element methods lead to state-space models with state dimension orders of $10^{5}$ or even $10^{6}$. Such large-scale models are impractical for real-time prediction or control. Finally, it is often the case that the parameters describing the STOP models are not accurately known or there are other model uncertainties. Consequently, it is often necessary to directly estimate the models from the experimentally collected data. All these facts call for the development of data-driven estimation and model validation approaches capable of estimating low-dimensional STOP models. This paper aims at developing and testing such approaches.

In our previous work~\cite{haber2020modeling}, we investigated the potential of using a subspace system identification method~\cite{verhaegen2007filtering,haber2020modelingHaberVerhaegen,haber2014subspace,haber2013identification} for estimating STOP models of refractive optical systems. In~\cite{haber2020modeling}, we considered a test case consisting of a single lens with an optomechanical support structure. By using the simulation data, we demonstrated that the subspace system identification method has a promising potential for accurately estimating low-order transient STOP models. However, the feasibility of the subspace identification method for estimating low-dimensional STOP models of reflective optics has not been investigated. Then, in~\cite{haber2021modeling}, we derived and experimentally verified a model of transient thermal dynamics of an 8-inch aluminum mirror prototype. In the same paper, we used model-order reduction techniques to develop low-order state-space models of thermal dynamics. The results reported in~\cite{haber2021modeling} indicate that the transient thermal dynamics of reflective optics can be approximated by low-order models. However, in~\cite{haber2021modeling}, we only consider thermal dynamics without coupling the heat equation with other equations mathematically describing thermal deformation and optical ray propagation. Consequently, it is not clear if an integrated STOP transient dynamics of reflective optics can be approximated by low-dimensional models.

Motivated by the promising results presented in~\cite{haber2020modeling,haber2021modeling}, and above described open research questions, in this paper, we investigate the feasibility and performance of the subspace system identification method for estimating STOP models of reflective optical systems. As a test case, we use a Newtonian telescope structure. We estimate and validate a state-space model of a transient STOP dynamics. First, we model the system in COMSOL Multiphysics. Then, we use LiveLink for MATLAB software module to export the wavefront aberrations data from COMSOL to MATLAB. This data is used to test the subspace identification method that is implemented in Python. Our results show that the large-dimensional STOP dynamics of the considered optical system can be accurately approximated by a low-dimensional state-space model. Due to its low-dimensional nature and state-space form, the estimated model can effectively be used for the prediction, estimation, and control of thermally-induced wavefront aberrations. Furthermore, the used estimation and validation procedures can be used for the development of feedforward adaptive optics compensation methods~\cite{haber2022dual,Roddier1999,tyson2010principles,Haber:13,Bonora2006,haber2021general,vogel2010modeling,polo2013linear}. The developed MATLAB, COMSOL, and Python codes are available online~\cite{stopCodesHaber2022,stopSubspaceCodesHaber2022}.

A few comments about the synergistic approach presented in this paper are in order. Since the purpose of this paper is to test the feasibility of the subspace identification method, we use simulated STOP data to test the identification approach. The next development stage is to experimentally verify the presented approach. This is a future research direction. In our accompanying article~\cite{haberMLSTOP2022}, we test the potential of using machine learning techniques for estimating low-order STOP models of the Newtonian telescope structure. The system identification approach presented in this paper and the machine learning approach presented in the accompanying paper~\cite{haberMLSTOP2022} complement each other. 

This paper is organized as follows. In Section~\ref{sec:systemSTOPmodel}, we present the STOP model and perform step response analysis. In Section~\ref{sec:systemIdentification}, we present the system identification approach and results. Finally, in Section~\ref{sec:conclusions}, we present conclusions and briefly discuss future research directions.

\section{SYSTEM STOP MODEL}
\label{sec:systemSTOPmodel}

\begin{figure}[H]
\centering 
\includegraphics[scale=0.9,trim=0mm 0mm 0mm 0mm ,clip=true]{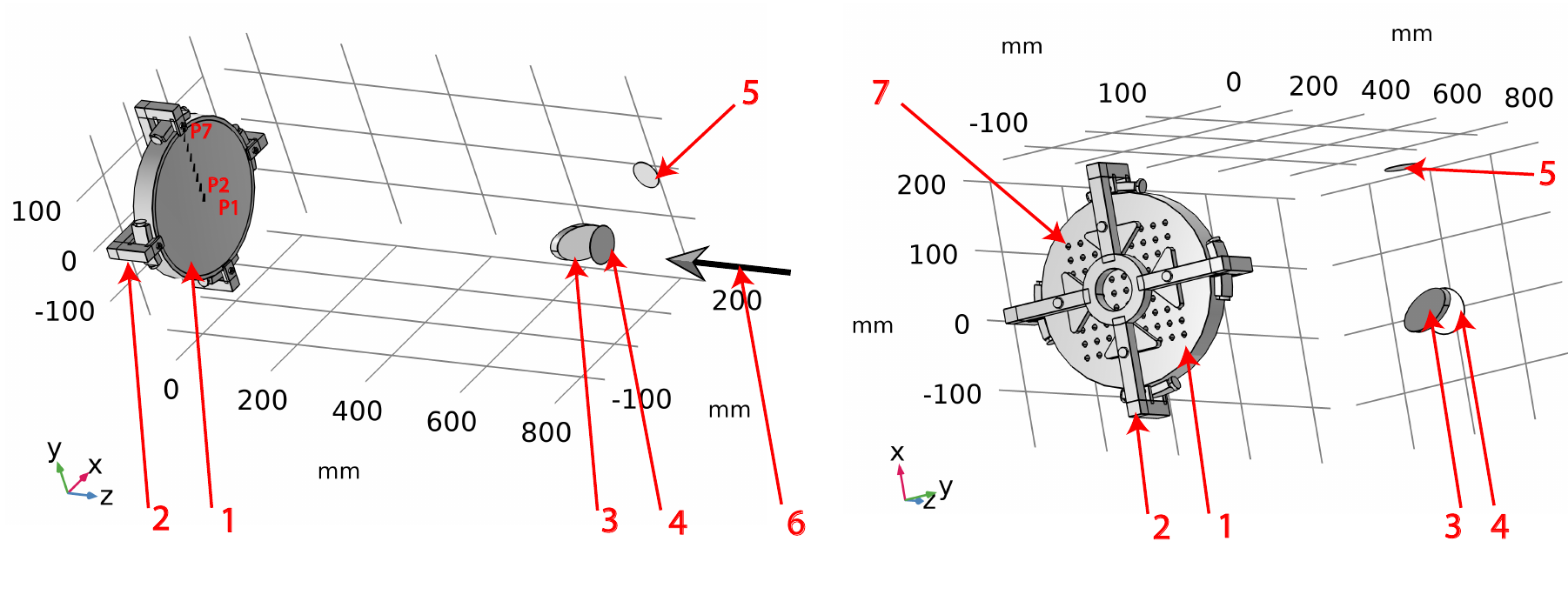}
\caption{Newtonian telescope structure. (1) Primary mirror. (2) Primary mirror support structure. (3) Secondary mirror. (4) Ray propagation obstruction. (5) Focal (image) plane. (6) Direction of rays entering the telescope. (7) 9 by 9 grid of holes on the mirror back side that are used to place heaters/coolers or thermocouples.}
\label{fig:Graph1}
\end{figure}

In this section, we develop the system STOP model. Figure~\ref{fig:Graph1} shows the system structure. This is a conceptual design obtained by combining a Newtonian telescope structure with a primary mirror support. We use ray-tracing parameters and dimensions from~\cite{comsolNewtonian2022} to perform a ray tracing analysis in COMSOL Multiphysics. Table~1 summarizes the most important geometrical and ray tracing parameters. The primary mirror, denoted by 1 in Fig.~\ref{fig:Graph1}, has holes on the back side that are used to place cooler/heater devices and thermocouples for observing the temperature. Holes, denoted by 7, are distributed over a 9 by 9 grid. The motivation for introducing the heaters/coolers originates from our previous work on designing feedback temperature control systems for optical components~\cite{haber2020modeling,haber2021modeling}. In our STOP simulations, heaters (heat inputs) are used to provide the heat power that increases the primary mirror and support structure temperatures, and consequently, introduces wavefront aberrations. Also, in our STOP simulations, we introduce external heat-flux disturbances acting on one side of the primary mirror. We are interested in developing a STOP model that relates the time series of the applied heat inputs and external heat-flux disturbances with the time-series of observed wavefront aberrations expressed in the Zernike basis. 

 The support structure of the primary mirror is denoted by 2 in Fig.~\ref{fig:Graph1}. A more detailed view of the mirror support structure is shown in Fig.~\ref{fig:Graph3}(b). We assume that the primary mirror and the support structure are made of an aluminum alloy, with the thermal and structural parameters given in Table~\ref{symbolGlossaryOptical}. Our modeling approach can easily be generalized to other mirror materials, mirror geometries, and mount structures.  Although we followed some guidelines~\cite{lockwoodOptics2022} for designing and modeling the support structure, the support mirror structure is not optimized from the structural and thermo-mechanical perspectives. The purpose of this paper is not to propose an optimized support structure, instead, the purpose of the paper is to test the ability of subspace identification techniques to estimate STOP models. The geometry of the primary mirror support structure does not have a significant influence on our estimation results. The secondary mirror is denoted by 3 in Fig.~\ref{fig:Graph1}. The ray propagation obstruction is denoted by 4. The image (focal plane) is denoted by 5. Arrow 6 denotes the direction of rays entering the telescope.
 
\begin{table}[H]
\centering
\label{symbolGlossaryOptical}
\begin{tabular}{|l| l |}
       \hline
        Entrance pupil diameter   & $0.25\;\; [m]$   \\
       \hline
        Primary mirror focal length   & $1\;\; [m]$   \\
       \hline
 Primary mirror conic constant & $-1$     \\
     \hline
 Primary mirror focal ratio & $4$     \\
       \hline
        Image plane position (relative to optical axis) & $0.2 [m]$     \\
       \hline
             Secondary mirror diameter  & $0.05 \;[m]$     \\
       \hline
             Secondary mirror offset  (relative to optical axis) & $0.0044194\; [m]$     \\
             
       \hline
              Image plane diameter  & $0.05\;[m]$     \\
       \hline
               Number of extra azimuthal points & $50$     \\
       \hline
               Primary mirror surface diameter & $0.26\; [m]$     \\
       \hline
		Primary mirror full diameter                      & $0.275\; [m] $       \\
	       \hline
			Primary mirror thickness                      & $0.035\; [m] $       \\
	       \hline
	       	Secondary mirror thickness                      & $0.01\; [m] $       \\
	       \hline
	       	       	Wavelength                       & $550 \; [nm] $       \\
	       \hline
	             	       	Mirror emissivity                       & $0.1 $       \\
	       \hline
	       	             	       	Ambient temperature                       & $293.15 \; [K] $       \\
		 \hline
			  	             	       	Heat transfer coefficient - convection                       & $5 \; [W/(m^{2}\cdot K) ]$   \\
		\hline
					  	             	       	Heat capacity at constant pressure                       & $900 \; [J /(kg\cdot K) ]$   \\
		 \hline

					  	             	       	Thermal conductivity                       & $238 \; [W /(m \cdot K) ]$   \\
		 \hline
		 					  	             	       	Coefficient of thermal expansion                    & $23 \cdot 10^{-6} \; [1/K]$   \\
		 \hline
		 					  	             	       	Density                    & $2700 \; [kg/m^{3}]$   \\
		 \hline
		 		 					  	             	       	Young's modulus                    & $70 \cdot 10^{9} \; [Pa]$   \\
		 \hline
		 		 					  	             	       	Poisson's ratio                    & $0.33$   \\
		 \hline

 \end{tabular}
 \caption{Optical, thermal, and structural parameters that are used to model the STOP system.} 
 \end{table}

\begin{figure}[H]
\centering 
\includegraphics[scale=0.8,trim=0mm 0mm 0mm 0mm ,clip=true]{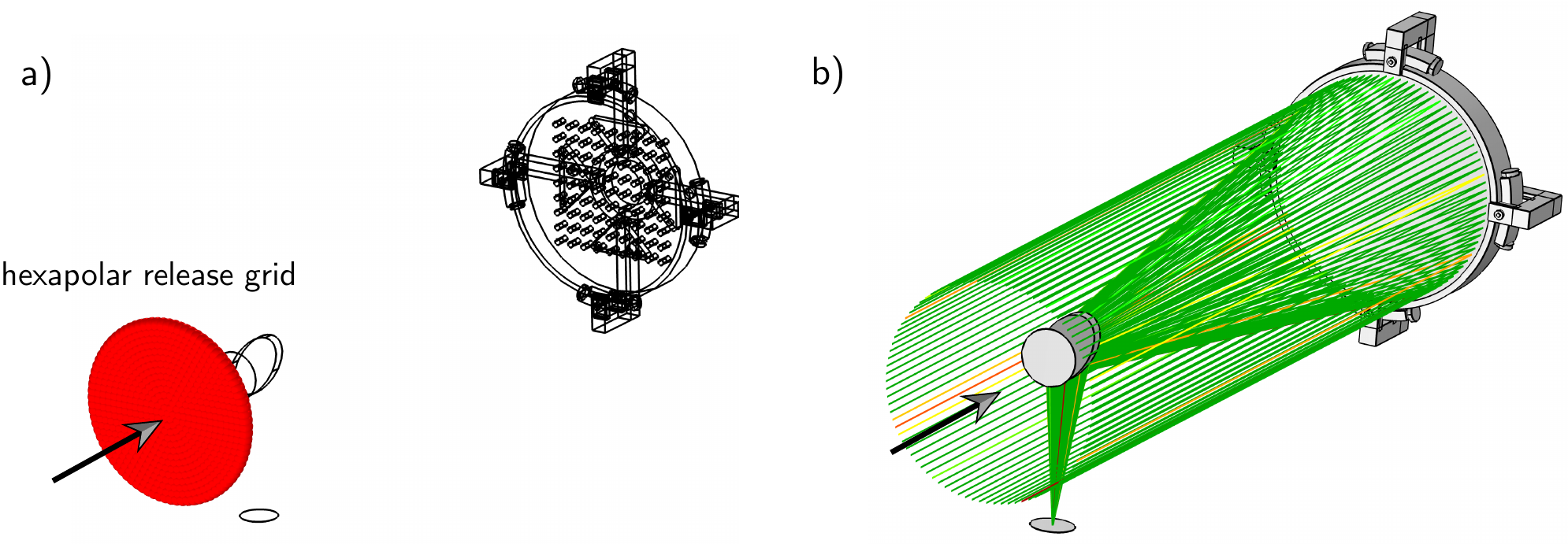}
\caption{(a) Ray release grid for ray tracing. (b) Simulated ray trajectories. Note that due to clarity, we only show a small portion of ray trajectories.}
\label{fig:Graph2}
\end{figure}

Figure~\ref{fig:Graph2}(a) shows a ray release grid and Fig.~\ref{fig:Graph2}(b) shows some of the simulated ray trajectories. For clarity, we do not overload Fig.~\ref{fig:Graph2}(b) with too many rays. That is, we only show ray trajectories of a small portion of the released rays.
 
\subsection{Control heater STOP results}

\begin{figure}[H]
\centering 
\includegraphics[scale=0.7,trim=0mm 0mm 0mm 0mm ,clip=true]{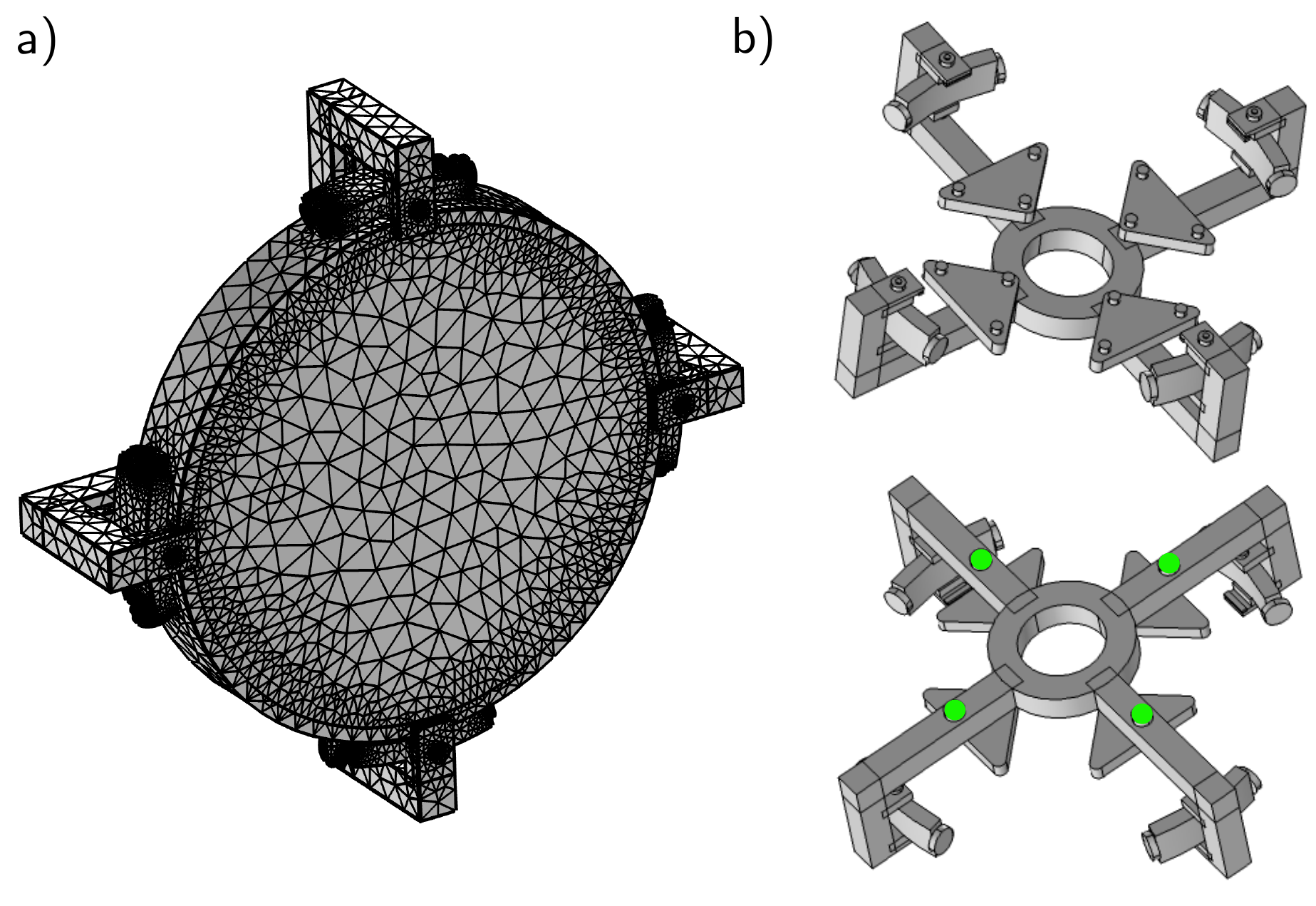}
\caption{(a) Meshed mirror and support structure CAD model. (b) Mirror support structure. The green marks denote the fixed constraints (the displacement is set to zero) for performing the STOP analysis.}
\label{fig:Graph3}
\end{figure}

We simulate a step response of the system, where the inputs are control heaters. We assume that the central and four neighboring heaters placed in the holes on the back of the primary mirror are active. We assume that every heater generates $4\;[W]$ of power. The heater power is constant during simulations. To perform the STOP analysis, we couple Geometrical Optics, Solid Mechanics, and Heat Transfer in Solids COMSOL Multiphysics modules. We first define and run a COMSOL study consisting of Solid Mechanics and Heat Transfer in Solids modules. This simulation run produces time-dependent temperature and displacement fields. Then, the results of this simulation are used in ray tracing simulations. To perform ray tracing simulations, we use the Geometrical Optics module. The COMSOL and MATLAB codes used to perform STOP analysis are posted online~\cite{stopCodesHaber2022}. Fig.~\ref{fig:Graph3}(a) shows the meshed geometry. The mesh contains around $109\cdot 10^{3}$ elements. Fig.~\ref{fig:Graph3}(b) shows the support structure of the primary mirror. The green marks denote the fixed constraints (the displacement is set to zero) for performing the STOP analysis.

\begin{figure}[H]
\centering 
\includegraphics[scale=0.7,trim=0mm 0mm 0mm 0mm ,clip=true]{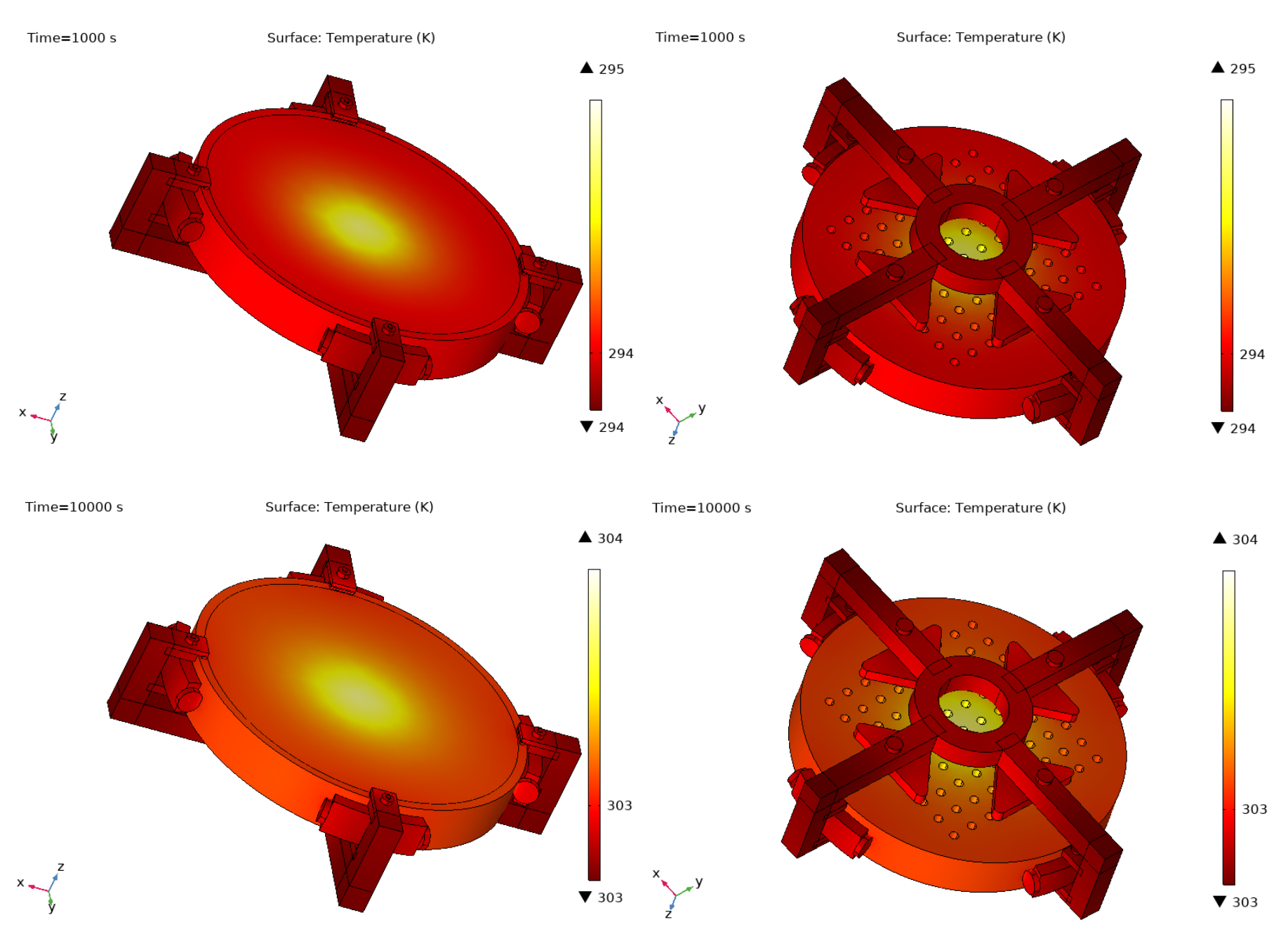}
\caption{STOP results for control inputs. Simulated temperature distribution at $10^{3}$ and $10^{4}$ seconds.}
\label{fig:Graph4}
\end{figure}

\begin{figure}[H]
\centering 
\includegraphics[scale=0.7,trim=0mm 0mm 0mm 0mm ,clip=true]{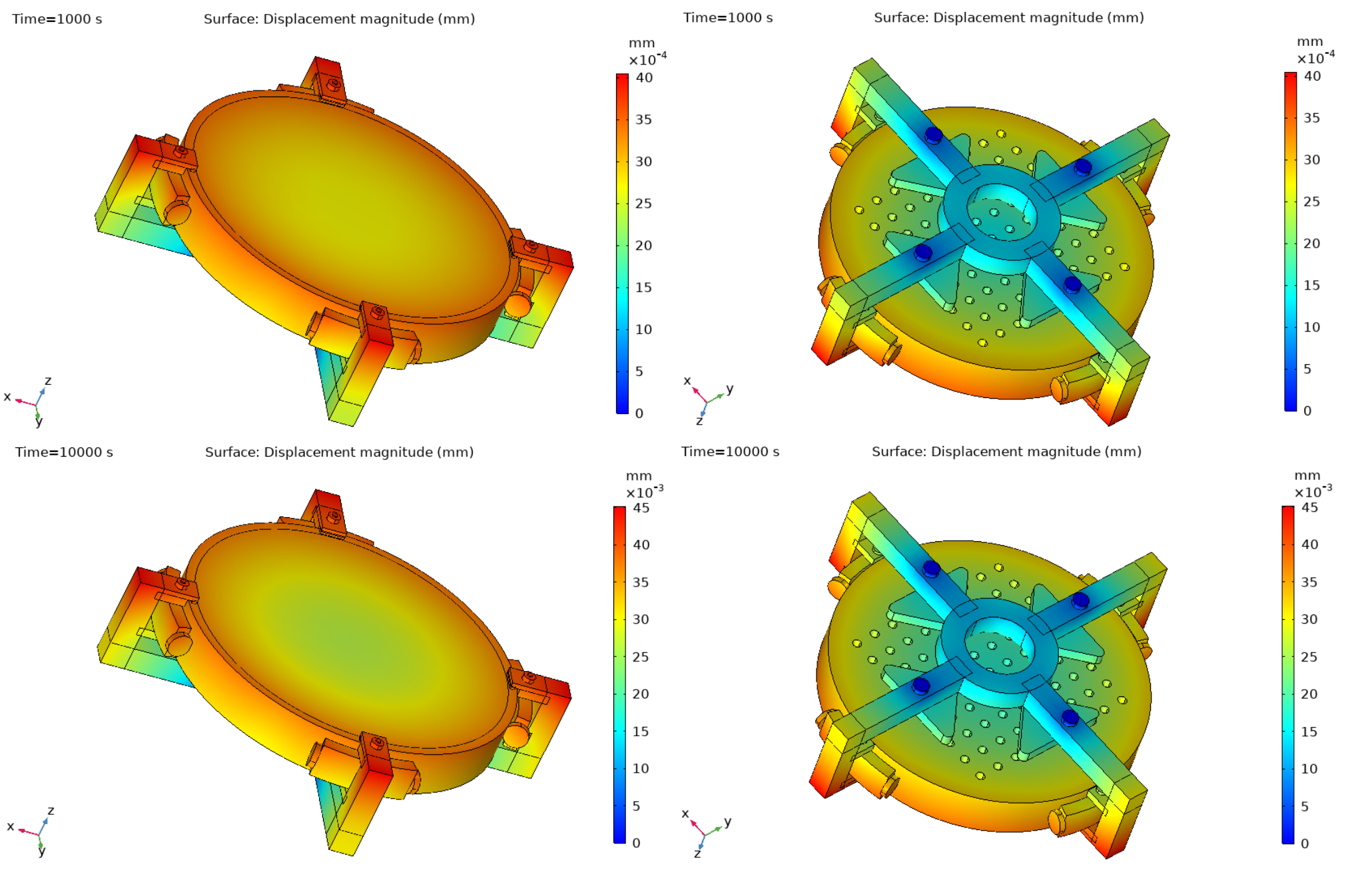}
\caption{STOP results for control inputs. Simulated displacement distribution at $10^{3}$ and  $10^{4}$ seconds.}
\label{fig:Graph5}
\end{figure}

\begin{figure}[H]
\centering 
\includegraphics[scale=0.8,trim=0mm 0mm 0mm 0mm ,clip=true]{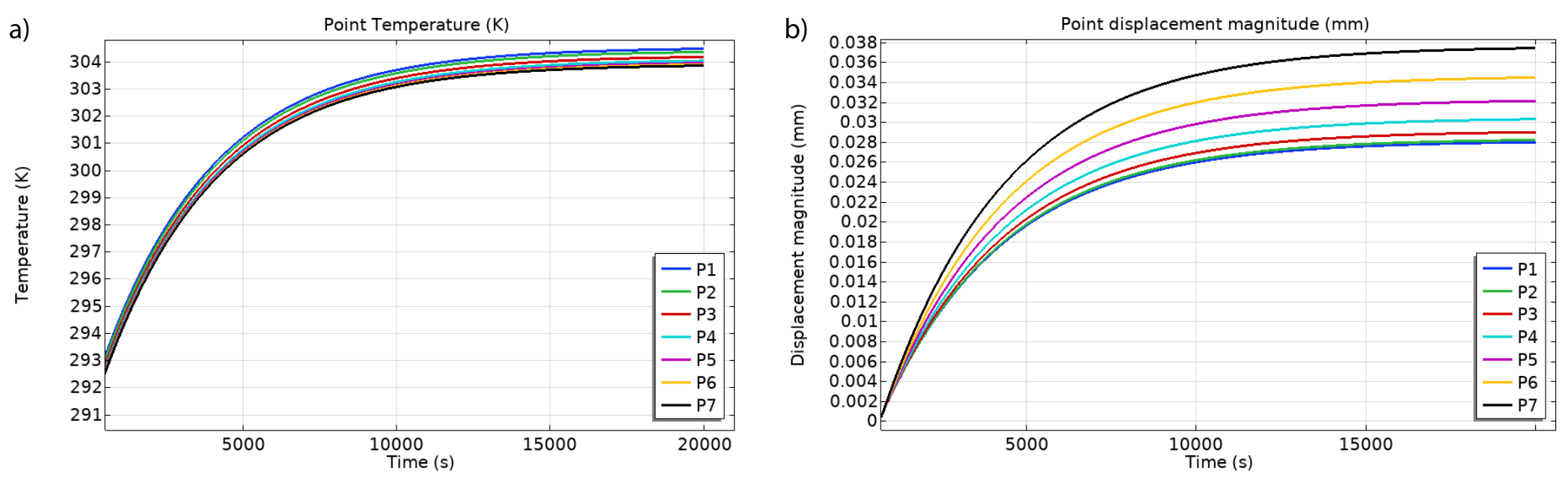}
\caption{STOP results for control inputs. (a) Temperature and (b) displacement transient responses at points P1, P2, ..., P7. The locations of these points are shown in Fig.~\ref{fig:Graph1}.}
\label{fig:Graph6}
\end{figure}

We simulate the transient STOP dynamics for $2\cdot 10^{4}$ seconds with a step size of $100$ seconds. The simulated temperature distributions are shown in Fig.~\ref{fig:Graph4} for $10^{3}$ and $10^{4}$ seconds. The simulated displacement distributions are shown in Fig.~\ref{fig:Graph5} for $10^{3}$ and $10^{4}$ seconds. Figure~\ref{fig:Graph6} shows (a) transient temperature and (b) displacement responses at the spatial locations defined by points P1, P2, ..., P7 that are shown in Fig.~\ref{fig:Graph1}. This graph can be used to estimate transient response parameters, such as rise time, settling time, and time constants. From Figs.~\ref{fig:Graph4} and \ref{fig:Graph5}, we can observe that the simulated temperature and displacement fields spatially correlate with the locations of the heat inputs. On the other hand, from Fig.~\ref{fig:Graph6}, we can observe that there are spatial gradients of temperature and displacement fields at the top surface of the primary mirror. At first look, the magnitudes of these gradients do not seem significant. However, wavefront aberrations results that are presented in the sequel, reveal that even these moderate gradients can cause significant wavefront aberrations and spot-diagram divergences in the focal plane.

\begin{figure}[H]
\centering 
\includegraphics[scale=0.7,trim=0mm 0mm 0mm 0mm ,clip=true]{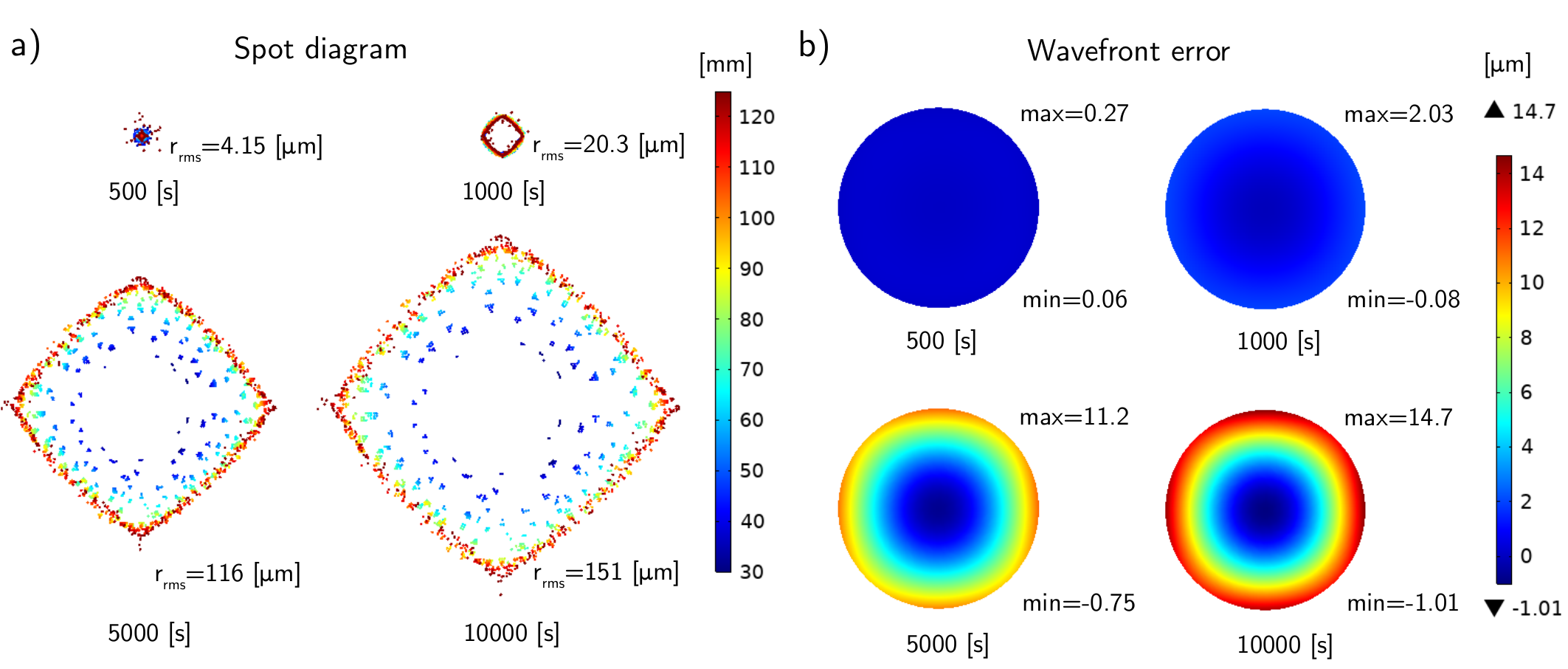}
\caption{STOP results for control inputs. (a) Spot diagrams and (b) wavefront error at the focal (image) plane at the time instants $500$, $10^{3}$, $5\cdot 10^{3}$, and $10^{4}$ seconds. }
\label{fig:Graph7}
\end{figure}

Next, we present spot diagrams and wavefront aberrations at the focal (image) plane. Fig.~\ref{fig:Graph7}(a) shows the spot diagrams at time instants $500$, $10^{3}$, $5\cdot 10^{3}$, and $10^{4}$ seconds. Fig.~\ref{fig:Graph7}(b) shows the wavefront aberrations at the same time instants. 

\subsection{Heat disturbance STOP results}

\begin{figure}[H]
	\centering 
	\includegraphics[scale=0.4,trim=0mm 0mm 0mm 0mm ,clip=true]{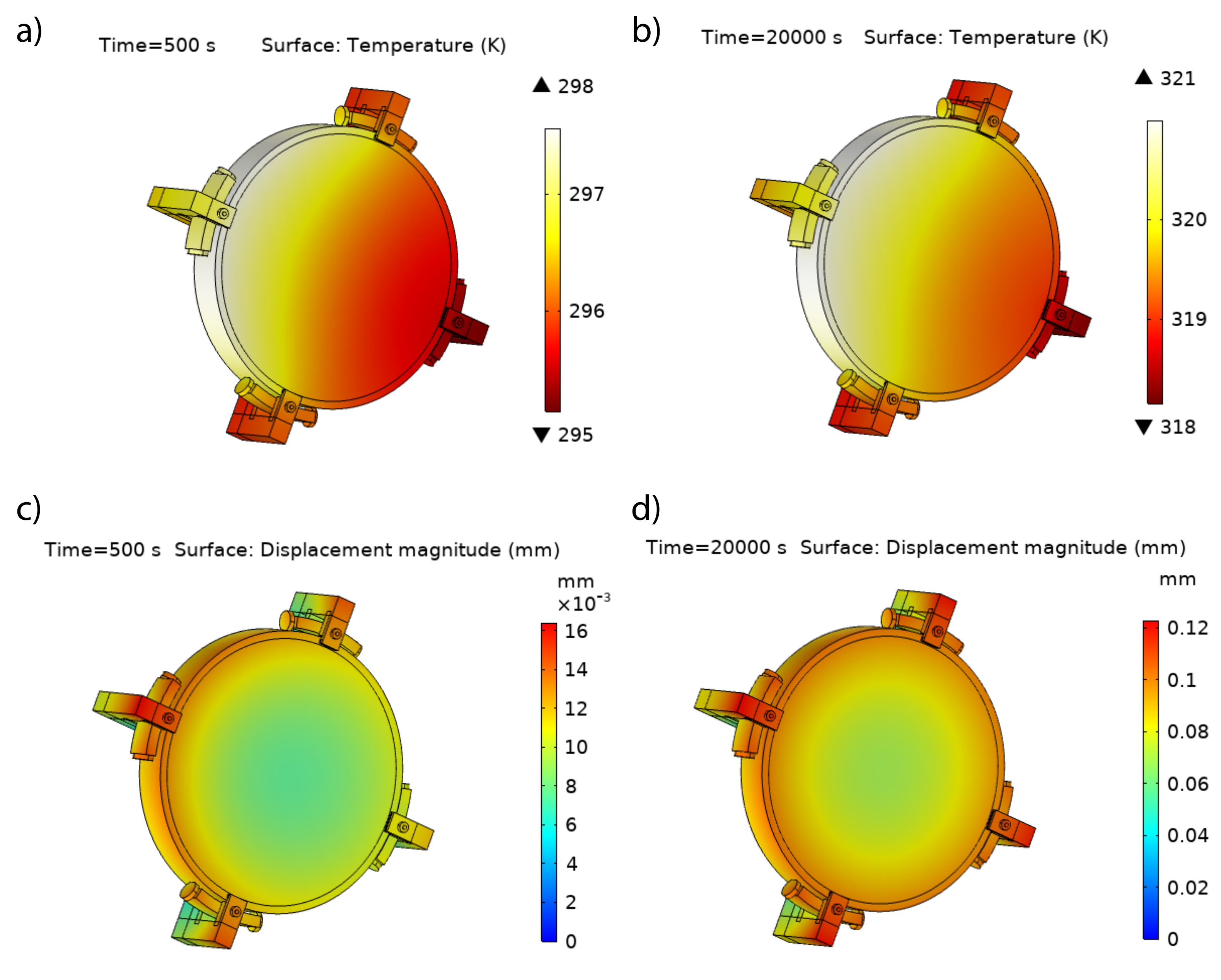}
	\caption{STOP results for external disturbance inputs. Simulated temperature spatial distribution at the time instant (a) $5 \cdot 10^2$ and (b) $10^4$ seconds. Simulated displacement spatial distribution at the time instant (c) $5 \cdot 10^2$ and (d) $10^4$ seconds.}
	\label{fig:Graph2disturbance}
\end{figure}

\begin{figure}[H]. 
	\centering 
	\includegraphics[scale=0.4,trim=0mm 0mm 0mm 0mm ,clip=true]{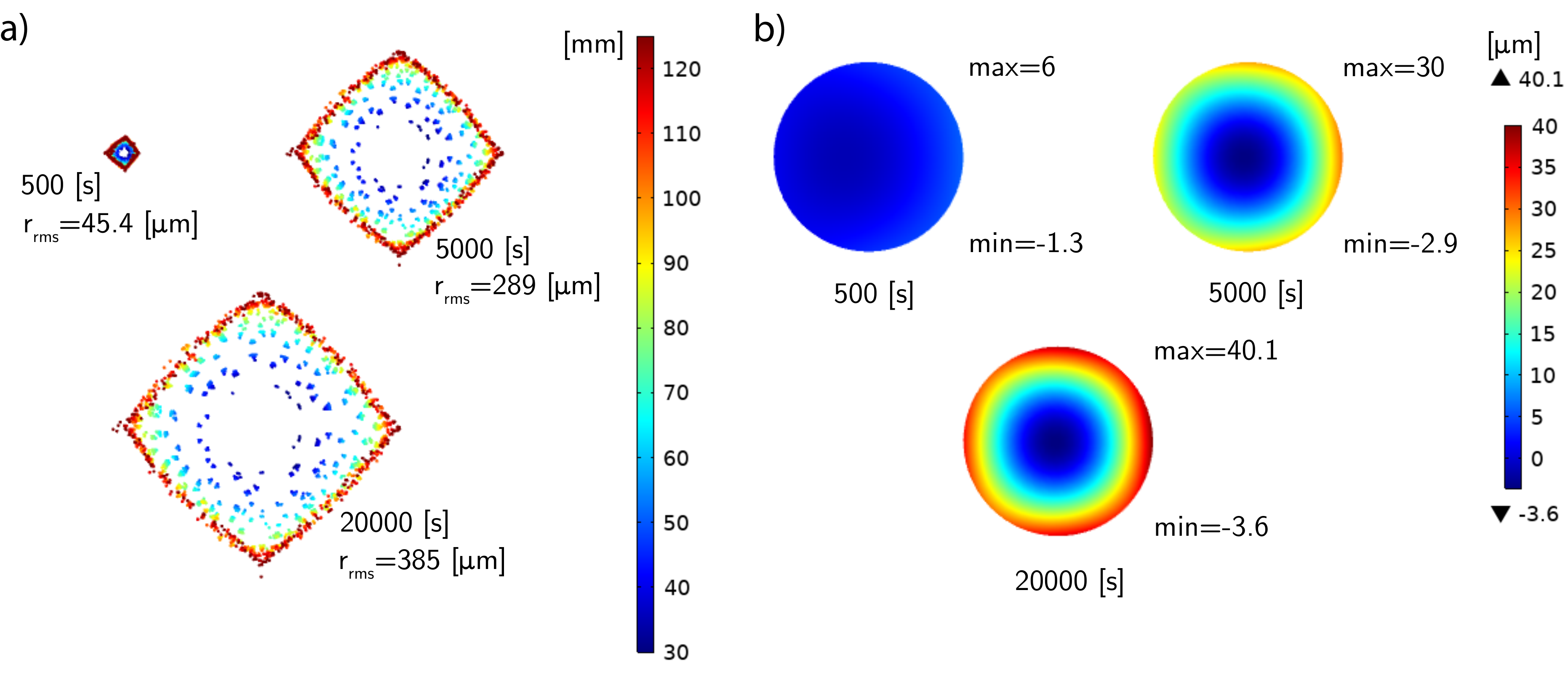}
	\caption{STOP results for external disturbance inputs. (a) Spot diagrams calculated at the image (focal) plane at the time instants $5\cdot 10^2$, $5\cdot 10^3$, and $2\cdot 10^4$ seconds.  The color legend quantifies the ray radial position (in the ray release plane) relative to the average over rays for each release feature (COMSOL's variable ``gop.rrel''). (b) Wavefront aberrations at the image (focal) plane at the identical time instants.}
	\label{fig:Graph3disturbance}
\end{figure}

Here, we present step response results where the inputs are the external heat-flux disturbances acting on one side of the mirror. The spatial location of the external disturbances is shown in Fig.~\ref{fig:Graph8} (light blue). We simulate the STOP dynamics for $2\cdot 10^{4}$ seconds with a step size of 100 seconds. In our STOP simulations, the total heat flux power of the external disturbances is $50\;[W]$. Figure~\ref{fig:Graph2disturbance}(a) and (b) show simulated temperature spatial distributions of the primary mirror at time instants $5 \cdot 10^2$ and $2 \cdot 10^4$ seconds. The panels (c) and (d) in the same figure show simulated displacement spatial distributions at identical time instants.
Figure~\ref{fig:Graph3disturbance}(a) shows the calculated spot diagrams at the image plane at the time instants $5\cdot 10^{2}$, $5\cdot 10^{3}$, and $2\cdot 10^{4}$ seconds. Figure~\ref{fig:Graph3disturbance}(b) shows the calculated wavefront aberrations at identical time instants. 

\section{SYSTEM IDENTIFICATION}
\label{sec:systemIdentification}

\begin{figure}[H]
\centering 
\includegraphics[scale=0.5,trim=0mm 0mm 0mm 0mm, clip=true]{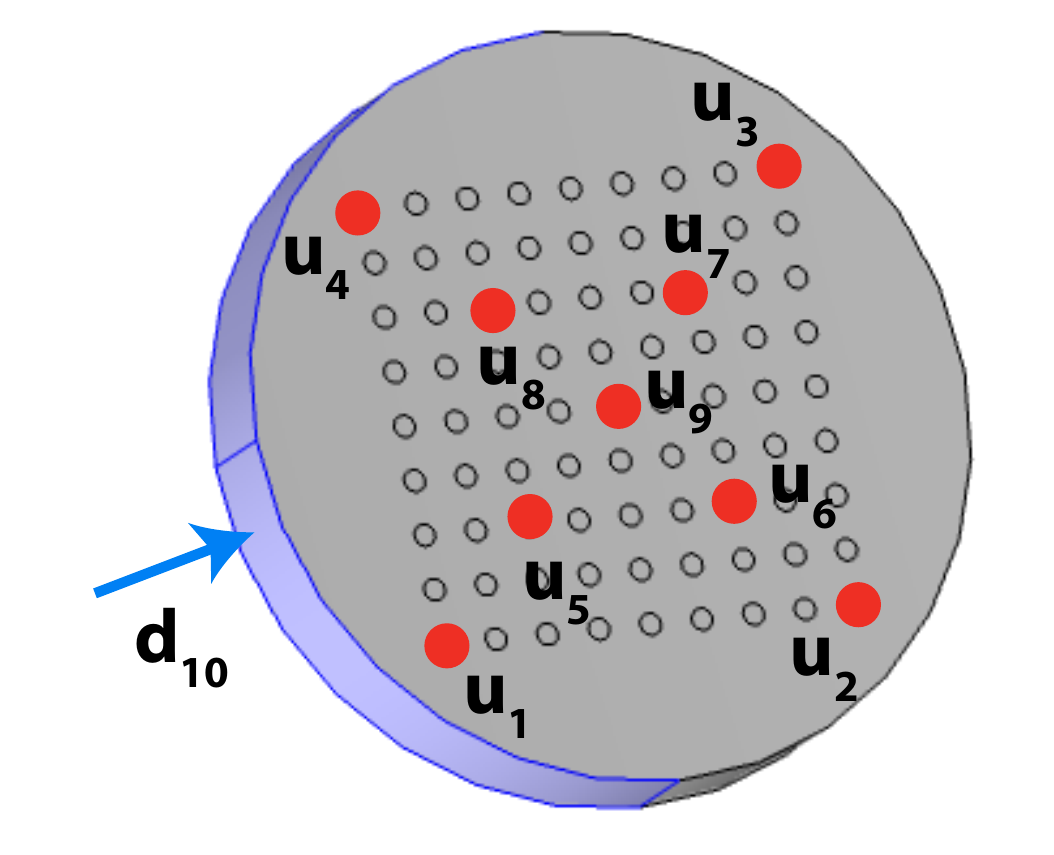}
\caption{Red circles denote the locations of control inputs $u_{1},u_{2},\ldots, u_{9}$ (heater locations) and blue area denotes the area on which external heat disturbance $d_{10}$ is acting upon.}
\label{fig:Graph8}
\end{figure}

In this section, we briefly summarize the used system identification method and present estimation and model validation results. Additional technical details related to the used subspace identification method can be found in~\cite{haber2019subspace,haber2019identification,haber2020modelingHaberVerhaegen}. Our goal is to estimate the following state-space model:
\begin{align}
\mathbf{x}_{k+1}& =A\mathbf{x}_{k}+B\mathbf{z}_{k} \label{ssModel1} \\
\mathbf{y}_{k} & = C\mathbf{x}_{k} \label{ssModel2}
\end{align}

where $\mathbf{x}_{k}\in \mathbb{R}^{n}$ is the system state vector, the subscript $k=0,1,2,\ldots$ of all vectors denotes a discrete-time instant, $\mathbf{z}_{k} \in \mathbb{R}^{10}$ is the input vector (consisting of control inputs and disturbances) at the discrete-time instant $k$, $\mathbf{y}_{k}\in \mathbb{R}^{r}$ is the vector consisting of selected Zernike coefficients, and $A\in \mathbb{R}^{n\times n}$, $B\in \mathbb{R}^{n\times 10}$, and $C\in \mathbb{R}^{r\times n}$ are the system matrices. The input vector $\mathbf{z}_{k}$ consists of control inputs $u_{1,k},u_{2,k},\ldots, u_{9,k}\in \mathbb{R}$ and the external heat disturbance $d_{10,k}\in \mathbb{R}$:
\begin{align}
\mathbf{z}_{k}=\begin{bmatrix}u_{1,k} & u_{2,k} & \ldots & u_{9,k} & d_{10,k} \end{bmatrix}^{T}
\end{align}
The control inputs $u_{1,k},u_{2,k},\ldots, u_{9,k}$ represent the heat power generated by the heaters, and external heat disturbance represents the power of the external heat flux acting on the mirror. Figure~\ref{fig:Graph8} shows the physical locations of the control inputs (red circles) and the external heat disturbance (blue area on the mirror side). 
The identification problem is to estimate the state order $n$ and state-space matrices $A,B$ and $C$ of the state-space model \eqref{ssModel1}-\eqref{ssModel2} by using time series of the collected input-output data $\{\mathbf{y}_{k},\mathbf{z}_{k} \}^{k=0,1,\ldots, N}$. That is, by using time-series of the collected Zernike coefficients, and by using time series of the control inputs and the external heat disturbance, we want to estimate the model order and state-space matrices.

We use a version of the subspace identification method that is derived and summarized in our previous papers~\cite{haber2019subspace,haber2019identification,haber2020modelingHaberVerhaegen,haber2020modeling}. We implemented the subspace identification method in Python. We used LiveLink for MATLAB module to generate data sets for testing the subspace identification method. The Python codes together with LiveLink for MATLAB and COMSOL Multiphysics codes are provided online~\cite{stopSubspaceCodesHaber2022}. In the sequel, we briefly describe the identification steps and present the results.

\subsection*{Step 1: Generate identification and validation data sets}

In this paper, we generate input-output data sets for testing the subspace identification method by simulating the system STOP model. This is a usual practice when developing and testing estimation approaches. Namely, in the development phase, the performance of the approach is first tested on simulation data. Once this initial testing phase is completed and the estimation approach is iteratively perfected, the next phase is to test the approach on experimentally collected data. In this paper, we are focused on testing the identification method by using the simulation data, and experimental verification of the developed approach is a future research direction.

\begin{figure}[H]
\centering 
\includegraphics[scale=0.38,trim=0mm 0mm 0mm 0mm, clip=true]{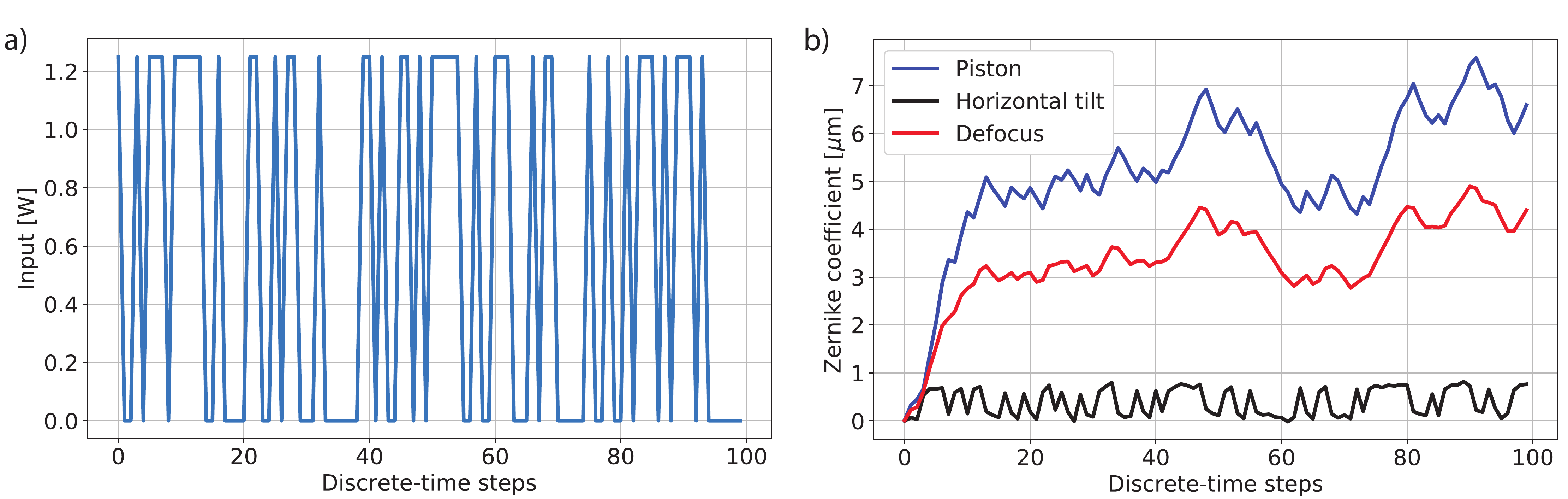}
\caption{(a) Example of the generated input sequence that is used for system identification. (b) Three dominant Zernike modes (coefficients) produced by the generated input sequences.}
\label{fig:Graph9}
\end{figure}

First, we generate input sequences for system identification. Generally speaking, input sequences have to be sufficiently rich such that they excite the system modes in the desired frequency range.  We generate input sequences (control inputs and the external heat disturbance) as binary pseudo-random numbers drawn from a uniform discrete distribution. We use the MATLAB function randi() to generate these sequences. Figure~\ref{fig:Graph9}(a) shows an example of the generated input sequence.
 
The input sequences are applied to the STOP model. By simulating this model, we obtain the output data. In total, by simulating the STOP model, we obtained the time series of the first $21$ Zernike modes (coefficients) that represent the output data. However, there are only 3 dominant Zernike modes whose magnitudes are significantly larger than the magnitudes of other modes. These modes are piston, horizontal tilt, and defocus, and they are shown in Fig.~\ref{fig:Graph9}(b). Consequently, we only include these modes in the output vector. That is, in our case, $r=3$ ($r$ is the dimension of the output vector).

We generate two independent sets of inputs. The first input set is used to generate the model outputs that are used for system identification. This data set is called the identification data set. The second input set which is statistically independent of the first input set is used to generate the outputs that are used for model validation. This data set is called the validation data set. To generate both of these data sets, we simulate the STOP model for the total time duration of $9 \cdot 10^{3}$ seconds with the discretization step of $300$ seconds. This gives in total $301$ data samples. The main issue with generating larger data sets is that it takes a significant amount of time to simulate the STOP model. We performed simulations on a desktop computer with 64 GB RAM and an Intel i9-10900 CPU. The generation of one data set on this computer takes at least 6 hours of computation time for a moderate-sized discretization mesh. For denser meshes, it might take several days to obtain a single data set.

\subsection*{Step 2: Estimation of the state-space model} 

\begin{figure}[H]
	\centering 
	\includegraphics[scale=0.35,trim=0mm 0mm 0mm 0mm, clip=true]{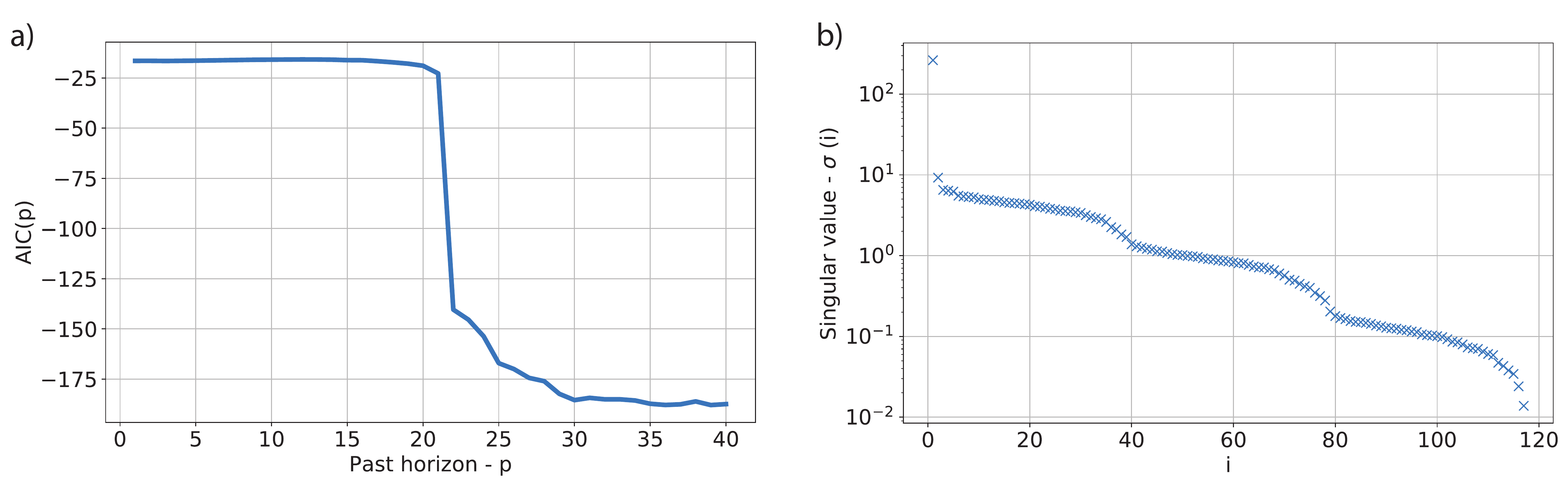}
	\caption{(a) AIC value as a function of the past window $p$. (b) Singular values of the data matrix used to estimate the model order.}
	\label{fig:Graph10}
\end{figure}

 First, we estimate a Vector AutoRegressive eXogenous (VARX) model. The VARX model is postulated such that the output of the system at the time instant $k$ is a linear combination of the past inputs and outputs, from $k-1$ until $k-p$, for more details, see~\cite{haber2019subspace,haber2019identification,haber2020modelingHaberVerhaegen,haber2020modeling}. Here $p$ is referred to as the past window. The first step of the subspace identification method is to estimate the parameters of the VARX model and past window $p$. We use a simple least-squares technique to estimate the VARX model parameters. We use the Akaike Information Criterion (AIC) to estimate the value of the past window $p$~\cite{lutkepohl2005new}. Figure~\ref{fig:Graph10}(a) shows the AIC as a function of the past window $p$. We select the past window $p=39$ that produces the smallest AIC value.

We use the estimated VARX model, and input-output identification data to form a data matrix. We estimate the state-sequence of the system by performing a singular value decomposition of this matrix~\cite{verhaegen2007filtering}. After the state sequence is estimated, we estimate the state-space matrices by solving a least-squares problem. During the state and state-space matrices estimation steps, we also estimate the state order $n$. The state order is estimated by detecting gaps in the plot of singular values. Namely, the state order can be estimated as a singular value index immediately after significant gaps. Figure~\ref{fig:Graph10}(b) shows the singular values. We can observe that the candidates for the state order are $n=2,3,6$ and higher state orders, such as for example $n=35$.

\subsection*{Step 3: Model validation and quality check}

\begin{figure}[H]
\centering 
\includegraphics[scale=0.32,trim=0mm 0mm 0mm 0mm, clip=true]{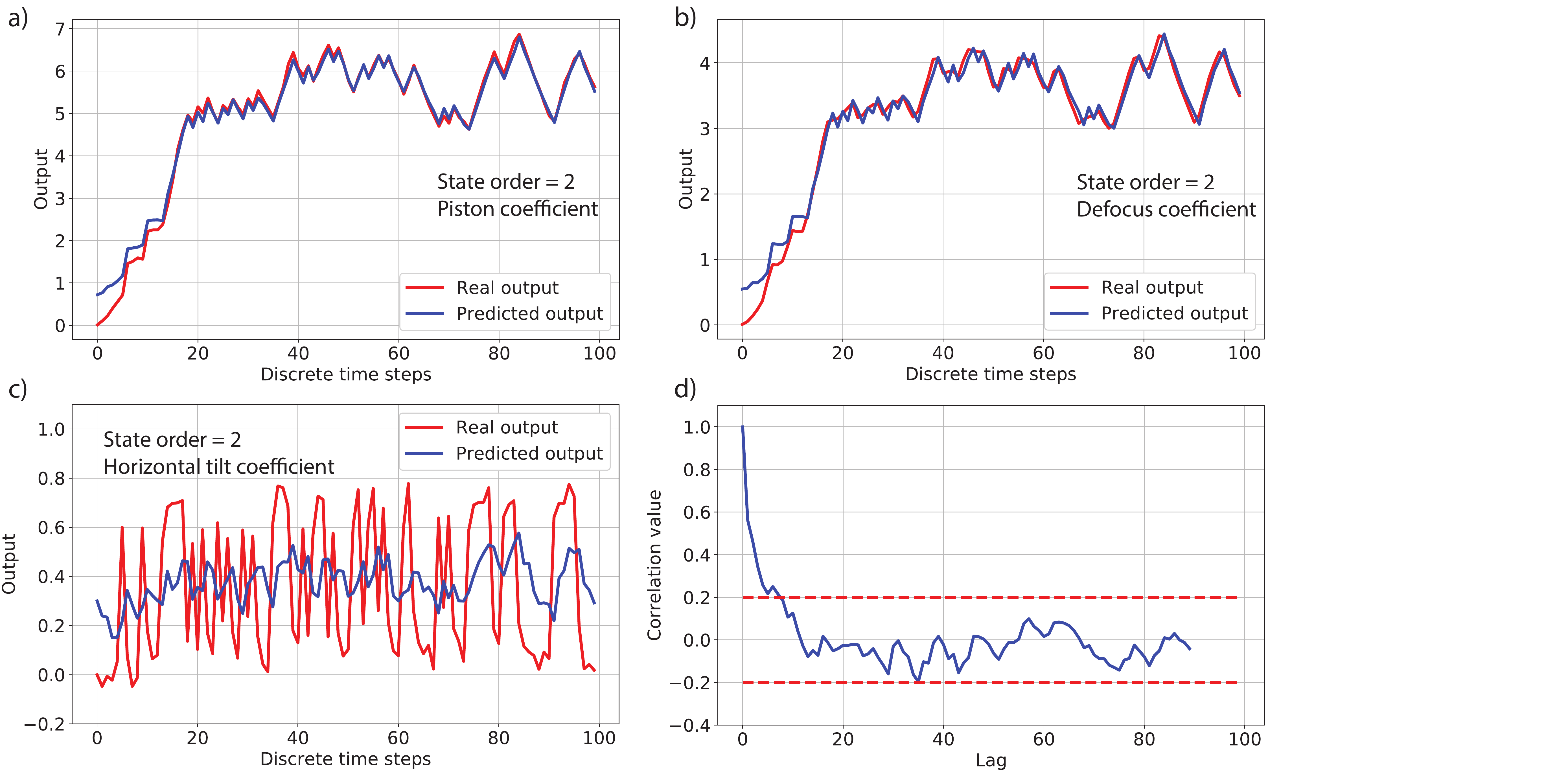}
\caption{Predicted and real Zernike modes: (a) piston, (b) defocus, and (c) horizontal tilt. (d) Correlation values of the defocus validation error. The red dashed lines are the bounds of the interval for testing the white-noise hypothesis. The results are generated for $p=39$, future window $f=5$, and state order of $n=2$. The relative validation error and Variance Accounted For (VAF) values are $5.99$ and $99.6$, respectively.}
\label{fig:Graph11}
\end{figure}

\begin{figure}[H]
\centering 
\includegraphics[scale=0.32,trim=0mm 0mm 0mm 0mm, clip=true]{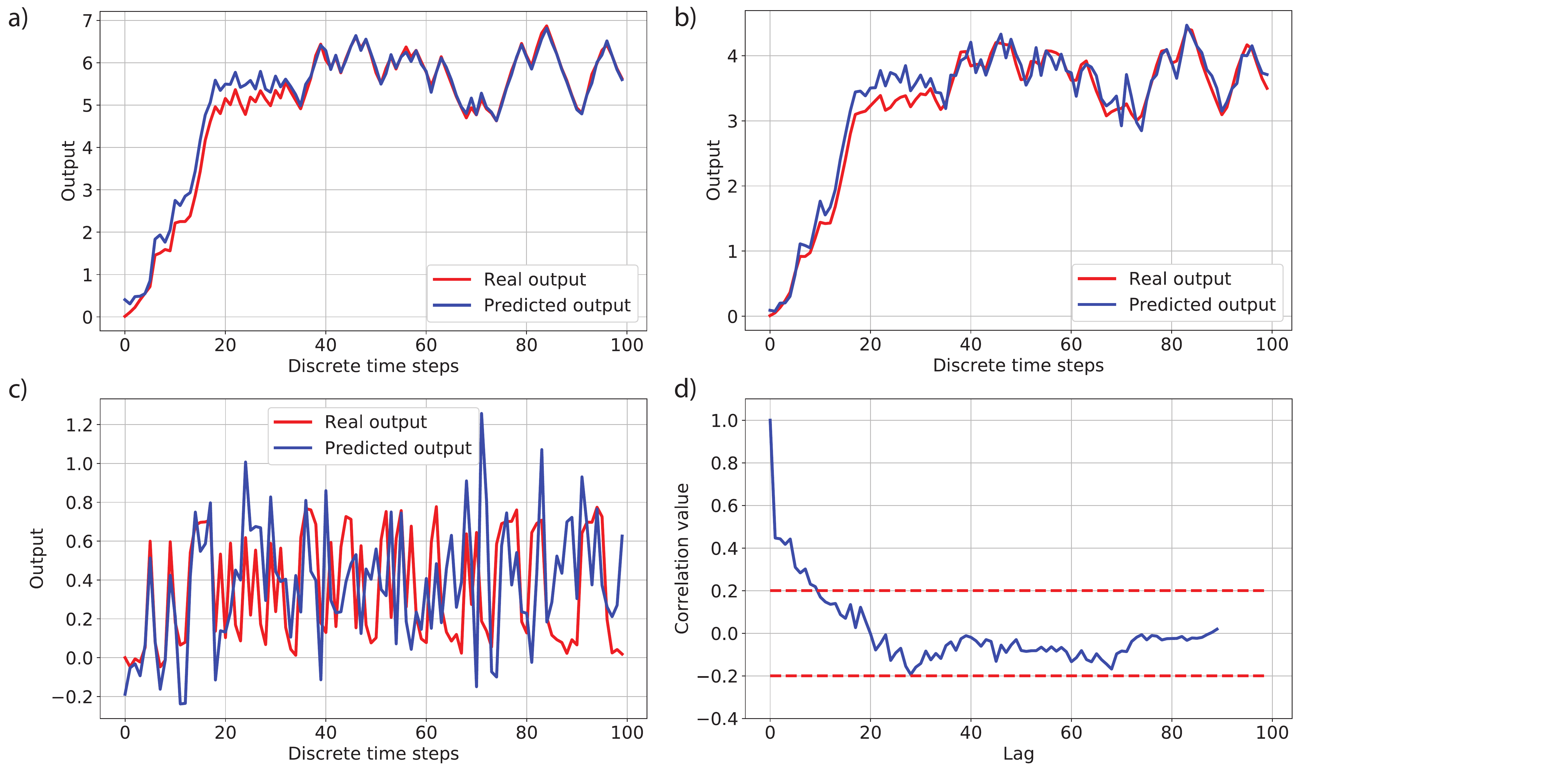}
\caption{The panel descriptions are identical to the panel descriptions in Fig.~\ref{fig:Graph11}. The results are obtained for $p=39$, $f=22$ and $n=35$. The relative validation error and VAF values are  $7.6$ and $99.42$, respectively.}
\label{fig:Graph12}
\end{figure}

Often, several models are estimated and the final model is chosen by validating and testing the performance of estimated models by using the validation data set. Following this practice, once we estimated the models for different state orders, we simulate models by using the input sequence from the validation data set. Then, the simulated model outputs are compared with the output from the validation data set. The error between the validation output (also called the real output) and the simulated output (also called the predicted output) is computed. This error is called the validation error or the model prediction error. The final model is selected as the model that produces the smallest validation error. Panels (a), (b), and (c) in Figure~\ref{fig:Graph11} show the real and predicted piston, defocus, and horizontal tilt Zernike coefficients (outputs), respectively, for the estimated state order $n=2$. Panels (a), (b), and (c) of Fig.~\ref{fig:Graph12} show the prediction and real value of the same coefficients determined for the estimated order of $n=35$.

On the other hand, panels (d) in these two figures show the correlation values of the validation error of predicting the defocus term. The red dashed lines represent the bounds of the interval for testing the white-noise hypothesis of the validation error. This interval is used to additionally validate the model quality. Ideally, if all information available in the data is captured by the final model, then the validation error should have a white noise property (this is also valid when the data is corrupted by the white measurement noise). If more than $95$ percent of correlation values are in the interval, then we can assume that the validation error has a white noise property. From the data in panels (d), we can observe that this is not the case since there is a strong correlation for smaller lag values. This implies that the estimation results can be improved by changing the model structure or choosing a different model order. However, from panels (a) and (b), we can observe that our estimated final models for both $n=2$ and $n=35$ are able to accurately predict the piston and defocus modes. However, the prediction is worse for the horizontal tilt mode. This is due to the fact that the horizontal tilt mode is more oscillatory and more difficult to be estimated. Furthermore, piston and defocus mode values are significantly larger than the values of the horizontal tilt mode. These results can be improved by some form of data scaling and detrending of the piston and defocus modes. The improvement of our simulation results is a future research direction.


\begin{figure}[H]
\centering 
\includegraphics[scale=0.38,trim=0mm 0mm 0mm 0mm, clip=true]{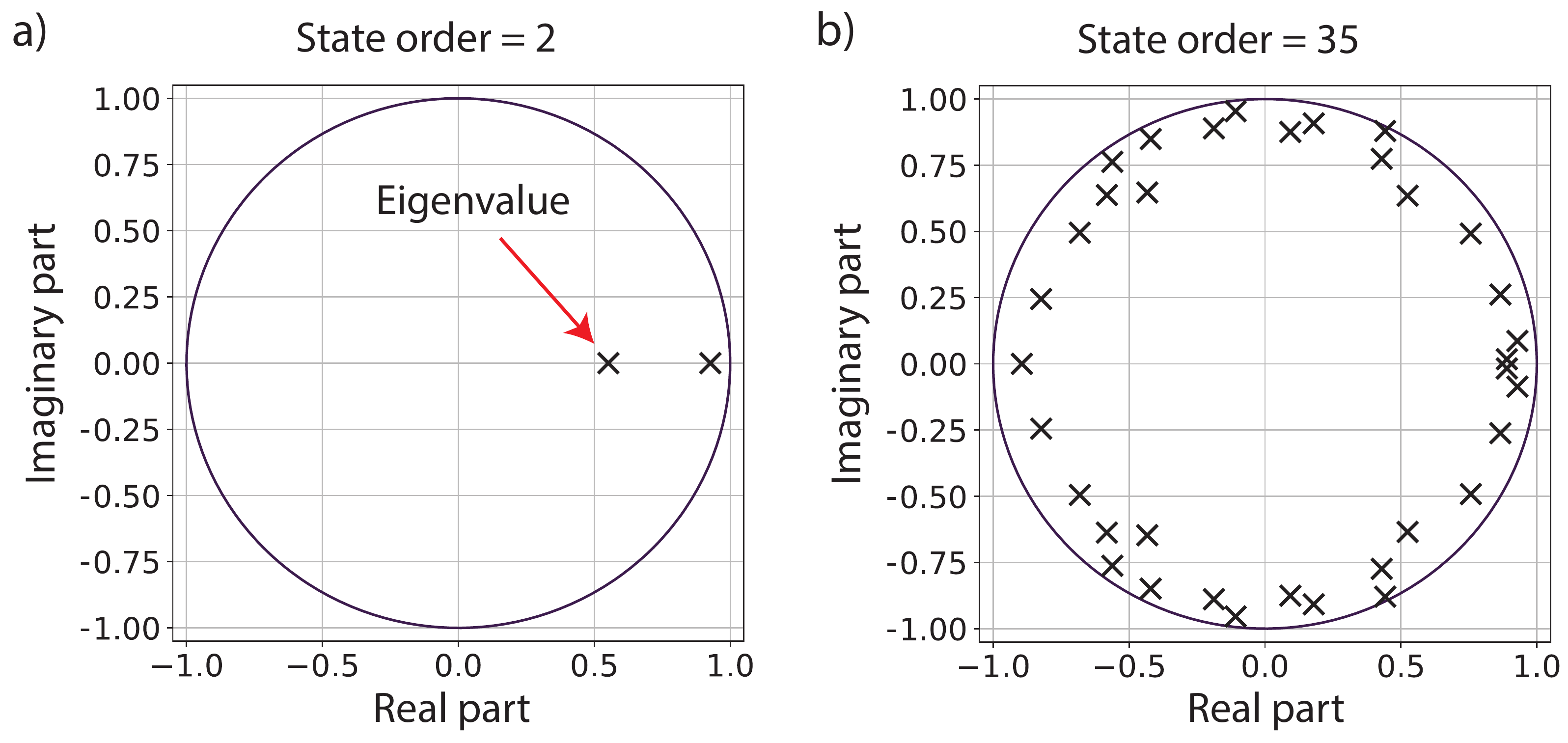}
\caption{Eigenvalues of the estimated model for (a) $n=2$ (other parameters given in caption of Fig.~\ref{fig:Graph11}) and (b) $n=35$ (other parameters given in caption of Fig.~\ref{fig:Graph12}).}
\label{fig:Graph13}
\end{figure}

As the final model quality check, we investigate the stability of the estimated models. Figure~\ref{fig:Graph13} shows the eigenvalues of the estimated models for (a) $n$=2 and (b) $n=35$. We can observe that both models are stable since all the eigenvalues are inside of the unit circle. Another important observation is that for $n=35$, the eigenvalues are clustered close to the unit circle. A thorough analysis of this phenomenon is left for future research. 

\section{Conclusion and Future Work}
\label{sec:conclusions}
In this paper, we investigated the feasibility of using the subspace identification method for estimating state-space models of transient Structural Thermal Optical Performance (STOP) dynamics of reflective optical systems. We tested the method on a Newtonian telescope structure. We obtained identification and test data sets by simulating the STOP model in COMSOL Multiphysics. Our results demonstrate that the subspace identification method is capable of estimating low-order STOP models of the dominant wavefront aberrations. Future research directions should be directed towards improving the estimation performance by proper data preprocessing and method tuning. Also, future research direction should be directed towards experimental verification of the subspace identification method.


\bibliography{sample} 
\bibliographystyle{spiebib} 

\end{document}